\documentclass[aps,reprint,prd,amsmath,showpacs,showkeys]{revtex4-1}

\usepackage{graphicx}
\graphicspath{{figures/}}

\begin{document}

\title{Spherical Gravitating Condensers in General Relativity}
\author{J. Bi\v c\'ak}
\author{N. G{\"u}rlebeck}
\affiliation{Institute of Theoretical Physics, Charles University, V Hole\v{s}ovi\v{c}k\'ach 2,\\180 00 Praha 8 - Hole\v{s}ovice, Czech Republic}
\affiliation{Max Planck Institute for Gravitational Physics, Albert Einstein Institute, Am M\"uhlenberg 1, D-14476 Golm, Germany}
\begin{abstract}
By a spherical gravitating condenser we mean two concentric charged shells made of perfect fluids restricted by the condition that the electric field is nonvanishing only between the shells. Flat space is assumed inside the inner shell. By using Israel's formalism we first analyze the general system of $N$ shells and then concentrate on the two-shell condensers. Energy conditions are taken into account; physically interesting cases are summarized in two tables, but also more exotic situations in which, for example, the inner shell may occur below the inner horizon of the corresponding Reissner-Nordstr{\"o}m geometry or the spacetime is curved only inside the condenser are considered. Classical limits are mentioned.
\end{abstract}

\maketitle

\section{Introduction}

Modeling physical systems by 2-dimensional thin shells sweeping out 3-dimensional timelike hypersurfaces in spacetime found numerous applications in general relativity and cosmology, in particular after the work of Israel \cite{Israel_1966} and his collaborators (see \cite{Barrabes_1991} for a more recent account). The material properties are characterized in terms of geometrical quantities like the jumps of the external curvature of the hypersurfaces. In contrast to pointlike or 1-dimensional sources this idealization is mathematically well defined \cite{Geroch_1987}.

In the following we shall be interested in shells made of a (2-dimensional) perfect fluid with a surface charge density. Israel's method was generalized to thin charged shells without pressure by de la Cruz and Israel \cite{Cruz_1967}. A comprehensive treatment of charged shells with pressure satisfying the polytropic equation of state was given by Kucha\v r \cite{Kuchar_1968}; Chase \cite{Chase_1970} placed no restriction on the equation of state and a spherically symmetric Reissner-Nordstr{\"o}m field inside the shell was admitted.

Until now a number of papers employed charged thin shells to tackle various problems, mostly under the assumption of spherical symmetry. For illustration: Boulware \cite{Boulware_1973} studied the time evolution of such shells and showed that their collapse can form a naked singularity if and only if the matter density is negative, in Ref. \cite{Farrugia_1979} the third law of black hole  mechanics was investigated by a charged shell collapsing in a Reissner-Nordstr{\"o}m field. Going over to most recent contributions (where many references to older literature can be found) , the shells are often renamed  ``membranes'' or ``bubbles;'' the equations of state become more exotic but the formalism remains. In Ref. \cite{Guendelman_2009} the stabilizing effects of an electric field inside a neutral shell made of dust or from a ``string gas'' (equation of state $p=-\frac 1 2 \sigma$) were studied using Israel's formalism, whereas in Ref. \cite{Belinski_2009} the authors analyze charged spherical membranes by the direct integration of the Einstein field equations with $\delta$-function sources (and show coincidences with the results of \cite{Israel_1966} and \cite{Chase_1970}). In particular, it is demonstrated in \cite{Belinski_2009} that acceptable parameters can be chosen such that stable charged membranes producing the over-extreme Reissner-Nordstr{\"o}m geometry with ``repulsive gravity'' effects exist. Gravity becomes repulsive also with uncharged ``tension shells'' if the tension is sufficiently high \cite{Katz_1991}.

From our perspective we wish to mention yet the work of King and Pfister \cite{King_2001} in which electromagnetic dragging (``Thirring'') effects are investigated by considering systems of two concentric spherical shells. Before (small) angular velocities are applied to the shells, the authors study the static two-shell model as we do in the present paper. However, in \cite{King_2001} for the purpose of dragging effects it was sufficient to assume a special system: the interior shell carries charge and tension but no rest mass, the exterior shell carries mass but no charge. No treatment of spherical condensers in the context of general relativity appears to have been given so far.

What, in fact, do we mean by such condensers? We consider two concentric spherical charged shells made from 2-dimensional perfect fluids with either pressure or tension, with arbitrary mass and charge densities. All parameters entering the problem are restricted by just one condition: the radial field may exist only in the region between the shells; of course, we take flat space inside the inner shell. We also regard energy conditions and allow the inner shell  to occur below the inner horizon of the Reissner-Nordstr{\"o}m geometry. The existence of the electric field between the shells is the basic feature of our definition of a condenser in general relativity. We shall see how our results corroborate various expressions for a quasilocal mass in Reissner-Nordstr{\"o}m spacetime. For a specific example of a plane gravitational condenser with a positive cosmological constant, see \cite{1974_Novotny}.

The gravitating condensers are primarily of theoretical interest. Since, however, their total charge vanishes, they may be closer to astrophysical models than objects carrying a net charge. Indeed, some numerical calculations \cite{Wilson_1975}
indicate that the gravitational collapse to neutron stars can lead to a charge separation but the whole system of star and envelope is neutral. A charge separation can also arise in plasma accreting into a black hole \cite{Ruffini_1975} (see also \cite{Ruffini_2000}). Recently, detailed numerical investigations of the collapse of a stellar core with a net charge were performed \cite{Ghezzi_2005,Ghezzi_2007} with understanding, however, that the total charge of the star may be zero.

Throughout the text geometrical units with $G=c=1$ are used.

\section{The classical system}
\begin{figure}[t]
\centering{
\includegraphics[scale=0.85]{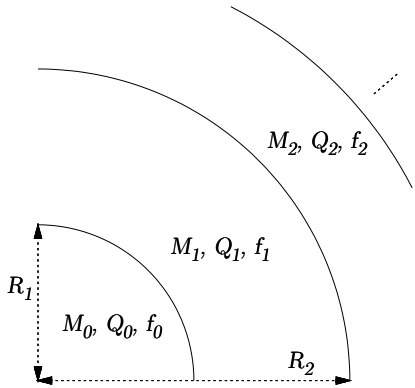}
\caption{\label{fig:structure} A sketch of the sequence of shells: $M_A$, $Q_A$ are the masses and charges in the respective spaces and $R_A$ the radii of the shells. The metric function $f_A$ will only play a role in Sec. \ref{sec:EinsteinMaxwell}.}}
\end{figure}

In order to introduce the notation and gain an intuition we now first analyze briefly a spherically symmetric condenser in Newton's gravity and Maxwell's electromagnetism in flat space. Let us consider charged perfect fluid spherical thin shells $\Sigma_{A}$ ($A=1,2,\ldots,N$) at radii $R_A$ ($R_A<R_{A+1}$) endowed with surface mass densities $\sigma_A$, surface charge densities $\eta_A$ and (2-dimensional) homogeneous surface pressures $p_A$.  $M_A$ ($A=0,\ldots,N$) denotes the total mass enclosed by a sphere with a radius $r\in(R_{A},R_{A+1})$ with $R_{0}=0$ and $R_{N+1}=\infty$. The charges $Q_{A}$ are defined analogously, see Fig. \ref{fig:structure}.
 After the solution for a general system with an arbitrary number of shells is obtained we discuss the case of a condenser ($N=2,~M_0=Q_0=Q_2=0,~Q_1=Q$) in full detail.
The gravitational potentials $U_A$ and the electrostatic potentials $\Phi_A$ ($A=0,\ldots,N$) in spherical coordinates read
\begin{equation}
\begin{split}
	U_A(r)&=-\frac{M_A}{r}+C_A,\\
		\Phi_A(r)&=\frac{Q_A}{ r}+D_A \quad\text{for}\,\, r\in [R_{A},R_{A+1}).
\end{split}
\end{equation}
The constants $C_A$ and $D_A$ are determined by the choice $D_{N}=C_{N}=0$  so that both $U_N$ and $\Phi_N$ vanish in infinity and by the requirement of continuity of the potentials across the shells. The surface densities can be determined from the jumps of the normal derivatives of the potentials across the shells  $\Sigma_A$. This yields ($A=1,\ldots,N$ here and in the remainder of this section)
\begin{equation}\label{Newton densities}
\begin{split}
\sigma_A=\frac{M_{A}-M_{A-1}}{4\pi R_A^2},\quad \eta_A=\frac{Q_{A}-Q_{A-1}}{4\pi R_A^2}.
\end{split}
\end{equation}

In order to obtain a stationary system the sum of all forces acting on a small element of each $\Sigma_A$ should vanish.
The forces acting on a surface element ${\mathrm d}S_A=R_A^2{\mathrm d}\Omega=R_A^2 \sin^2\theta{\mathrm d}\theta{\mathrm d}\varphi$ have only radial components because of the symmetry. The gravitational forces are given by \mbox{${\mathrm d}F_{(A)G}=-\frac{ (M_{A}^2-M_{A-1}^2)}{8\pi R_A^2}{\mathrm d}\Omega$}. These are always pointing inward for positive surface mass densities $\sigma_A$. The electrostatic contributions are \mbox{${\mathrm d}F_{(A)E}=\frac{(Q_{A}^2-Q_{A-1}^2)}{8\pi R_A^2}{\mathrm d}\Omega$}. The pressure forces ${\mathrm d}F_ {(A)p}$ on ${\mathrm d}S_A$ are given by $2p_AR_A{\mathrm d}\Omega$; these point always outwards for positive pressures $p_A$ and inwards for negative pressures (tension). 

The equilibrium is achieved if the total force on each ${\mathrm d}S_A$ vanishes:
\begin{equation}\label{Newton pressure}
\begin{split}
16\pi R_A^3p_A&=M_{A}^2-M_{A-1}^2-Q_{A}^2+Q_{A-1}^2.
\end{split}
\end{equation}
Equations \eqref{Newton densities} and \eqref{Newton pressure} give a complete solution with $3 N+2$ free parameters; e.g., $M_A,~Q_A$ can be chosen for $A=0,\ldots,N$ and $R_A$ for $A=1,\ldots,N$.

\section{\label{sec:EinsteinMaxwell}The Einstein-Maxwell system}

Now we turn to spherical condensers in the Einstein-Maxwell theory. First we consider a more general system of $N$ shells $\Sigma_A$ as in the Newtonian case, then we analyze thoroughly just the condensers ($N=2, M_0=Q_0=Q_2=0, Q_1=Q$). By $\Sigma_A$ we now understand $3$-dimensional timelike hypersurfaces representing the histories of the individual shells in the spacetime.

 Since the system is spherically symmetric and between the thin spherical shells there is electrovacuum, the spacetime must consist of $N+1$ pieces, $V_A$, $A=0,\ldots,N$, of Reissner-Nordstr{\"o}m spacetimes with the following line elements:
\begin{equation}
\begin{split}
	{\mathrm d}s_A^2=&-f_A(r_A){\mathrm d}t_A^2+\frac{1}{f_A(r_A)}{\mathrm d}r_A^2\\
	&+r_A^2({\mathrm d}\theta^2+\sin^2\theta{\mathrm d}\varphi^2),\\
	f_A(r_A)=&1-\frac{2M_A}{r_A}+\frac{Q_A^2}{r_A^2},
\end{split}
\end{equation}
where coordinates $x_{(A)}^\mu$ have ranges $t_A\in(-\infty,\infty)$, $r_A\in [R_{A-},R_{A+})$, $\theta\in [0,\pi]$, $\varphi\in[0, 2\pi)$ and the constants $M_A$ and $Q_A$ denote the mass and the charge parameters; we take $R_{1-}=0$ and $R_{N+}=\infty$. In general the coordinates $t_A$ jump across the shells (related to the jumps of the red shift factors due to the massive shells), but $R_{A+}$ and $R_{(A+1)-}$ must coincide since they have an invariant geometrical meaning giving the proper areas of the shells; so we set $R_{A+}=R_{(A+1)-}\equiv R_{A+1}$ and drop the index $A$ for the radial coordinate $r$. The angles $\theta$ and $\varphi$ change smoothly due to the symmetry.

In order to obtain timelike hypersurfaces $\Sigma_A$, it is assumed that $f_{A}(R_{A+1})>0$ and $f_{A+1}(R_{A+1})>0$, i.e., the shells are either situated in the exterior of the outer event horizon of the respective piece of the Reissner-Nordstr{\"o}m spacetime or inside the inner event horizon. 

On the hypersurfaces $\Sigma_A$ ``inner'' coordinates $(\xi^c_{(A)})=(\tau_A, \theta,\varphi)$ are chosen, $\tau_A$ is the proper time of an observer with fixed $(R_A,\theta,\varphi)$; ${\mathrm d}\tau_A=[f_A(R_A)]^{\frac 1 2}{\mathrm d}t_{A}=[f_{A+1}(R_A)]^{\frac 1 2}{\mathrm d}t_{A+1}$. The metric in $V_A$ is denoted by $g_{(A)\mu\nu}$.  Indices of tensor quantities on $\Sigma_A$ are labeled by $a,b,\ldots$. The metrics $h_{(A)ab}$ induced by $g_{(A-1)\mu\nu}$ and $g_{(A)\mu\nu}$ on $\Sigma_{A}$ from both sides read
 \begin{equation}
 h_{(A)ab}{\mathrm d}\xi_{(A)}^a{\mathrm d}\xi_{(A)}^b=-{\mathrm d}\tau^2_A+R_{A}^2({\mathrm d}\theta^2+\sin^2 \theta{\mathrm d}\varphi^2).
\end{equation}

The electric field is given by
\begin{equation}
 F_{(A)}^{\mu\nu}=\frac{Q_A}{r^2}(\delta^\mu_t\delta^\nu_r-\delta^\mu_r\delta^\nu_t)\,\,\, \text{for}\,\,\, r\in [R_{A},R_{A+1}).
\end{equation}
The hypersurfaces $\Sigma_A$ are endowed with surface mass densities $\sigma_A$, surface pressure densities $p_A$ and surface charge currents $s_{(A)}^a$.
The tangential electrical \emph{net} current on $\Sigma_A$ reads (see \cite{Kuchar_1968})
\begin{equation}
\begin{split}
	4\pi s_{(A)c}=& \left(F_{(A)\alpha\beta}\frac{\partial x_{(A)}^{\alpha}}{\partial \xi_{(A)}^c}n_{(A)}^{\beta}\right.\\
	&\left.-F_{(A-1)\alpha\beta}\frac{\partial x_{(A-1)}^{\alpha}}{\partial \xi_{(A-1)}^c}n_{(A-1)}^{\beta}\right)\Big|_{r=R_A}\\
	=&\frac{Q_{A-1}-Q_{A}}{ R_A^2}\delta_c^\tau,
\end{split}
\end{equation}
where $n_{(A-1)}^\alpha$ ($n_{(A)}^\alpha$) denotes the outwards pointing normal of $\Sigma_A$ as seen from $V_{A-1}$ ($V_{A}$). There can be more currents representing counter-rotating 
charges whose net contribution to the motion cancels out so that the spherical symmetry is preserved.

The extrinsic curvature of $\Sigma_A$ defined respectively by $x_{(A-1)}^\mu$, $g_{(A-1)\mu\nu}$ and $x_{(A)}^\mu$, $g_{(A)\mu\nu}$ is denoted by $K_{(A-1)cd}$ and $K_{(A)cd}$, its jump 
$K_{(A)cd}-K_{(A-1)cd}$ across $\Sigma_A$ by $[K_{(A)cd}]$, analogously for its trace $K_{A}=K_{(A)cd}h_{(A)}^{cd}$.
The surface stress-energy tensor $t_{(A)ab}$ on $\Sigma_A$ is now determined by \cite{Israel_1966,Barrabes_1991,Kuchar_1968}
\begin{equation}\label{eq:stress-energy-tensor general}
 	t_{(A)cd}=\frac{1}{8\pi}([K_{(A)cd}]-h_{(A)cd}[K_{A}]).
\end{equation}
In our case of spherical shells the jumps are given by
 \begin{equation}\label{jumps of the extrinsic curvature}
 \begin{split}
  [K_{(A)cd}]&=L_A(R_A){\mathrm{diag}}\Big(\frac{L'_A(R_A)}{L_A(R_A)},-R_A,-R_A\sin^2\theta\Big),\\
  [K_{A}]&=-L'_A(R_A)-\frac{2}{R_A}L_A(R_A),\\
  L_A(r)&=f_{A}^{\frac 1 2}(r)-f_{A-1}^{\frac 1 2}(r);
\end{split}
\end{equation}
here a prime denotes a derivative with respect to $r$.
In the case of a perfect fluid at rest its $4$-velocity is simply $u_{(A)}^c=\delta^c_\tau $ and the stress-energy tensor reads
 \begin{equation}\label{eq:stress-energy-tensor perfect fluid}
  t_{(A)cd}=(\sigma_A+p_A)u_{(A)c}u_{(A)d}+p_A h_{(A)cd}.
 \end{equation}
The surface mass density and the surface pressure density can be read off Eqs. \eqref{eq:stress-energy-tensor general}-\eqref{eq:stress-energy-tensor perfect fluid}:
 \begin{equation}\label{eq:mass and charge density}
 	\sigma_A=-\frac{1}{4\pi R_A}L_A(R_A),\quad p_A=\frac{1}{8\pi}L'_A(R_A)-\frac{\sigma_A}{2}.
 \end{equation}

We shall use extensively the last two relations. To get a physical insight, consider just one charged shell of radius $R$ with the Minkowski space inside and Reissner-Nordstr{\"o}m field with parameters $M,~Q$ outside. Equation \eqref{eq:mass and charge density} then implies
\begin{subequations}
\begin{align}
 \sigma&=\frac{1}{4\pi R}\left(1-\sqrt{1-\frac{2M}{R}+\frac{Q^2}{R^2}}\right),\label{eq:mass density one shell}\\
 p&=\frac{\overline{M}^2-Q^2}{16\pi R^2(R-\overline{M})},\label{eq:pressure one shell}
\end{align}
\end{subequations}
where $\overline{M}=4\pi R^2\sigma$ is the rest mass of the shell. The total energy of the shell at rest with a fixed charge $Q$ and a fixed pressure $p$ is given by $M$ and can be interpreted as function of $R$. The result \eqref{eq:pressure one shell} coincides precisely with Eq. (51) in \cite{Kuchar_1968}; this follows from the ``second'' equilibrium condition $M'(R)=0$, the total energy is extremal in equilibrium. The expression for $\sigma$ \eqref{eq:mass density one shell} follows directly from the ``first'' equilibrium condition, Eq. (50) in \cite{Kuchar_1968} (in which a typographical error must be corrected in the last term on the right-hand side), which represents conservation of the total energy of the shell at rest given by
\begin{equation}\label{eq:total energy}
 M(R)=\overline{M}+\frac{Q^2}{2R}-\frac{\overline{M}^2}{2R}. 
\end{equation}
The first term is the rest energy of the shell, the second describes the electromagnetic interaction energy of the particles of the shell, the last can be interpreted as the gravitational interaction energy --- see \cite{Kuchar_1968} for more details on the subtleties of such an interpretation.

\section{Condensers}

\subsection{Classical condensers}

In the case of a condenser ($N=2, M_0=Q_0=Q_2=0, Q_1=Q$) the Eqs. \eqref{Newton densities} and \eqref{Newton pressure} simplify to
\begin{subequations}\label{eq:classical results}
\begin{align}
 \sigma_1&=\frac{M_1}{4\pi R_1^2},&\qquad\sigma_2&=\frac{(M_2-M_1)}{4\pi R_2^2},\\
 \eta_1&=\frac{Q}{4\pi R_1^2},&\qquad\eta_2&=-\frac{Q}{4\pi R_2^2},
\end{align}
\begin{align}
\begin{split}
16\pi R_1^3 p_1 &=M_1^2-Q^2,\\
16\pi R_2^3 p_2 &=(M_2^2-M_1^2)+Q^2.
\end{split}
\end{align}
\end{subequations}
In this case the inner shell is not influenced by the field of the outer shell and must be in equilibrium in its own gravitational and electrical field. This can be achieved even with vanishing pressure if $M_1^2=Q^2$.
 However, the outer shell is always attracted by gravity and Coulomb force due to the inner shell. Therefore, the outer shell cannot consist of dust and $p_2$ must be positive to prevent the collapse.
Let us also remark that the limit to an electrostatic dipole shell ($R_1\to R_2$, $Q\to\infty$) implies that either the surface pressure density or the surface mass density is also unbound. In this case the net force acting on a volume element containing both shells $\Sigma_A$ has to be carefully treated to obtain physically meaningful results.

\subsection{Einstein-Maxwell condensers}

Let us now restrict to the case of a relativistic condenser. Primarily we wish to investigate whether the shells can be made of dust and whether the inner shell can be hidden below the horizon.  Since the spacetime outside of the exterior shell is part of a Schwarzschild spacetime the radius of the outer shell must satisfy $R_2>2M_2$, so no horizon can exist there. In the following we will measure all quantities in units of $M_1$ and denote them by small 
Latin letters or by a hat in the case of the pressure and the mass density, e.g., $R_A=M_1 r_A$ and $\sigma_A=\frac{\hat \sigma_A}{M_1}$. In the case of positive $M_1$ this notation does not bear the risk of any confusion; however, if $M_1\leq 0$ and $Q^2>0$, then $\sigma_1< 0$. We assume $M_1>0$ in the remainder. The stability of such shells was discussed in \cite{Kuchar_1968,Chase_1970}. The stability considerations apply for each shell entirely analogously in our case and so we do not repeat them here.

We obtain the densities and pressures from Eq. \eqref{eq:mass and charge density}:
\begin{subequations}
\begin{align}
 \hat\sigma_1=&\frac{1}{4\pi r_1^2}\left(r_1-\sqrt{r_1^2-2r_1+q^2}\right),\notag\\
 \hat s_{(1)}^c=&\frac{q}{4\pi r_1^2}\delta^c_\tau,\\
 \hat p_1=&\frac{1}{8\pi r_1}\left((r_1-1)(r_1^2-2r_1+q^2)^{-\frac 1 2}-1\right),\notag\\
\notag\\
\hat\sigma_2=&\frac{1}{4\pi r_2^2}\left(\sqrt{r_2^2-2r_2+q^2}-\sqrt{r_2^2-2m_1r_2}\right),\notag\\
\hat s_{(2)}^c=&-\frac{q}{4\pi r_1^2}\delta^c_\tau,\\
\hat p_2=&\frac{1}{8\pi r_2}\Big(\left(r_2-m_2 \right)\left(r_2^2-2m_2r_2 \right)^{-\frac 1 2}\notag\\
&-\left(r_2-1 \right)\left(r_2^2-2r_2+q^2 \right)^{-\frac 1 2} \Big).\notag
\end{align}
\end{subequations}
In the case of an under-extreme Reissner-Nordstr{\"o}m spacetime, $q^2<1$, the additional conditions $r_1<r_-=1-\sqrt{1-q^2}$ or $r_1>r_+=1+\sqrt{1-q^2}$ must be considered; the same must hold for $r_2$ because the shells can be at rest only in the regions where the Killing vector $\frac{\partial}{\partial t}$ is timelike. Furthermore, $r_2>2m_2$ must always be satisfied.

\subsubsection{The inner shell}

The energy conditions (dominant, weak, null, strong) and their implications are investigated below. Even though the calculations are in principle basic, they are tedious, so they are not shown here. Instead, the results are summarized in Table \ref{table:EnergyCondition:interior} in full detail. Looking solely at the inner shell yields the generic case of one charged, spherical thin shell and thus it is of interest of its own.
\begin{table*}[h!tb]
\begin{ruledtabular}
\begin{tabular}{llll}
\multicolumn{2}{l}{Energy condition}  & Range of $r_1$ & Range of $q^2$\\
\hline
Positive mass density & $\hat\sigma_1\geq 0$ & $r_1\in (0,2]$ & $q^2 \in (2 r_1-r_1^2,2 r_1]$\\
    & & $r_1\in(2,\infty)$ & $q^2 \in [0,2r_1]$\\
Null  &  $\hat\sigma_1+\hat p_1\geq 0$ & $r_1\in [\frac 8 9,1)$ & $q^2 \in [q_{2-}^2,q_{2+}^2]$\footnote{$q_{2\pm}=\frac{r_1}{8}\left(12-3r_1\pm\sqrt{9r_1^2-8r_1}\right)$}\\
  &  & $r_1\in [1,2]$ & $q^2 \in (2r_1-r_1^2,q_{2+}^2]$\\ 
	&  &  $r_1\in (2,\infty)$ &  $q^2 \in [0,q_{2+}^2] $\\ 
Weak & $\hat\sigma_1\geq 0$, $\hat\sigma_1+\hat p_1\geq 0$ & \multicolumn{2}{c}{the same as null}\\
Dominant & $\hat\sigma_1\geq |{\hat p_1}|$ & $r_1\in [\frac 8 9,1)$ & $q^2 \in [q_{2-}^2,q_{2+}^2] $\\
                &  &  $r_1=1$ & $q^2\in (1,\frac 5 4]$\\
								&	& $r_1\in (1,\frac {25} {12}]$ &  $q^2 \in \left[\frac{r_1}{8}\left(20-3r_1-3\sqrt{8r_1+r_1^2}\right),q_{2+}^2\right] $\\
								&  & $r_1\in (\frac {25} {12},\infty]$ & $q^2\in[	0,q_{2+}^2]$\\																									
Strong  &  $\hat \sigma_1+2\hat p_1\geq 0$,~$\hat \sigma_1+\hat p_1\geq 0$ & $r_1\in (1,2]$ & $q^2\in(2 r_1 - r_1^2,r_1]$ \\
                &  & $r_1\in(2,\infty)$ & $q^2\in[0,r_1]$
\end{tabular}

\end{ruledtabular}
\caption{\label{table:EnergyCondition:interior}Energy conditions for the inner shell. The table is to be read as follows: For example, the first two lines mean that the mass density $\sigma_1$ is positive if either $0<R_1\leq 2M_1$ and $2M_1R_1-R_1^2<Q^2\leq 2M_1R_1$ or if $R_1>2M_1$ and $0\leq Q^2\leq 2M_1 R_1$.}
\end {table*}

The null and the weak energy conditions are equivalent in this setting as shown in Table \ref{table:EnergyCondition:interior}. Therefore, we refer to the null energy condition only. From Table \ref{table:EnergyCondition:interior} it becomes clear that positive mass shells exist for all radii (though for a given charge of the shell there exists a minimal radius of $\frac {q^2}{2}$). In particular, $r_1<1$, $q^2<1$ can always be chosen; the shell then lies below the inner horizon. If, additionally, the second shell is chosen to be situated outside the outer horizon, which can always be achieved, the two shells of the condenser will be separated by two horizons. A massless shell is obtained for $q^2=2r_1$; this lies always below the inner horizon if there is one. 

If the energy conditions are to be satisfied, then the radius $r_1$ has to have a lower bound of $\frac 8 9$. However, as long as $r_1<1$ we have $q^2>1$. Thus, the spacetime between the two shells is a piece of an over-extreme Reissner-Nordstr{\"o}m spacetime and no horizons are present. If $r_1>1$, then $q^2<1$ is possible, but no horizon is present here either, since the inner shell lies outside of the outer horizon.  
Therefore, \emph{horizons can only be present if some energy condition is violated}. In particular the tension (negative pressure) has to be high if the inner shell is situated below the inner horizon. This resembles the problem considered by Novikov: the matter of a static charged sphere below the inner horizon of the Reissner-Nordstr{\"o}m geometry must have a high tension \cite{Novikov_1970}.
If additionally the charge $q$ is chosen arbitrarily close to zero, a lower bound for $r_1$ of $2$ for the null and stromg energy condition and of $\frac{25}{12}$ for the dominant energy condition is implied. 

The pressure $\hat p_1$ is always non-negative if the spacetime between the shells is a piece of an under-extreme Reissner-Nordstr{\"o}m spacetime and the inner shell is situated outside the outer horizon. In all other cases it is negative. Hence, a shell producing an over-extreme Reissner-Nordstr{\"o}m spacetime or a shell situated below the inner horizon can only be supported by tension. In the first case contrary to the second all energy conditions can be satisfied. 

If the inner shell is made of dust then -- similarly to the Newtonian case -- the spacetime between the two shells corresponds to a piece of an extremely charged Reissner-Nordstr{\"o}m spacetime, i.e., $|q|=1$ and the shell must be situated outside the horizon, $r_1>1$. The mass density evaluates in the extremely charged case to $\hat\sigma_1=\frac{1}{4\pi r_1^2}>0$. Since $\hat p_1=0$, all energy conditions (null, dominant and strong) are satisfied by such a shell. 

\subsubsection{The outer shell}

The situation for the exterior shell is more involved since more parameters have to be taken into account. Therefore, we treat in detail only the case when the inner shell is made of charged dust, i.e., $\hat p_1=0$ -- this is sufficient to exhibit some generic features of the condensers. As one easily checks the pressure $\hat p_2$ is always positive if $r_1>1$ in order to prevent the collapse. Thus, a positive mass density ensures the weak energy condition and vice versa; similarly, the null energy condition is equivalent to the strong energy condition. Consequences of all energy conditions are summarized in Table \ref{table:EnergyCondition:exterior}. It is seen that, \emph{a condenser can consist of an inner shell made of charged dust  and an exterior shell} ($r_2>r_1$), \emph{both shells satisfying all the energy conditions}. 

It is worth noting, that, in all discussed cases included in Table \ref{table:EnergyCondition:exterior} a mass parameter $M_2$ can be chosen smaller than $M_1$ for any radius $r_2$. This is because the electromagnetic field decreases the value of a (quasi) local mass energy at given $r$. Indeed, various results for quasilocal masses like those of Hawking, Penrose, Brown and  York, and others (though not Komar) lead to the same result for the mass energy inside a sphere of radius $r$ in Reissner-Nordstr{\"o}m spacetime
\begin{align}\label{eq:quasilocalmass}
 E(r)=M-\frac{Q^2}{2r},
\end{align} 
(see, e.g., \cite{Tod_1983,Bergquist_1992,Bose_1999}). This corresponds to the fact that the charge weakens the strength of the gravitational field (recall, for example, that the surface gravity of an extreme Reissner-Nordstr{\"o}m black hole vanishes). Let us also note that the formula \eqref{eq:quasilocalmass} does not contradict the conservation law \eqref{eq:total energy} since there $M(R)$ represents the total Schwarzschild mass outside the shell with radius $R$ and flat spacetime inside, whereas \eqref{eq:quasilocalmass} is the expression for the energy of the Reissner-Nordstr{\"o}m field inside the radius $r$ (cf. also \cite{Bose_1999}).
\begin{table*}[htb]
\begin{ruledtabular}
\begin{tabular}{llll}
Energy condition & & Range of $r_2$ & Range of $m_2$\\
\hline
Positive mass density & $\hat\sigma_2\geq 0$ & $r_2\in (1,\infty)$ & $m_2 \in [1-\frac{1}{2r_2},\frac{r_2}{2})$\\
Null  &  $\hat\sigma_2+\hat p_2\geq 0$ & $r_2\in (1,2]$ & $m_2 \in [m_{2,1+},\frac{r_2}{2})$\footnote{$m_{2,1\pm}=\frac{ 2}{9r_2}(-2 + 2 r_2 + r_2^2\pm \sqrt{4 - 8 r_2 + 9 r_2^2 - 5 r_2^3 + r_2^4})$}\\
   &  & $r_2\in (2,\infty)$ & $m_2 \in [m_{2,1-},\frac{r_2}{2}) $\\ 
Weak & $\hat\sigma_2\geq 0$, $\hat\sigma_2+\hat p_2\geq 0$ & \multicolumn{2}{c}{the same as positive mass density}\\
Dominant & $\hat\sigma_2\geq |{\hat p_2}|$ & $r_2\in [\frac{3+\sqrt{5}}{2},\infty)$ & $m_2 \in [m_{2,2-},m_{2,2+}] $\footnote{$m_{2,2\pm}=\frac{ 2}{25r_2}(-2 + 6 r_2 + 3r_2^2\pm \sqrt{4 - 24 r_2 + 49 r_2^2 - 39 r_2^3 + 9r_2^4})$}\\
Strong  &  $\hat \sigma_2+2\hat p_2\geq 0$,~$\hat \sigma_2+\hat p_2\geq 0$ & \multicolumn{2}{c}{the same as null}
\end{tabular}
\end{ruledtabular}
\caption{\label{table:EnergyCondition:exterior}Energy conditions for the exterior shell and ranges of $r_2$ and $m_2$.}
\end {table*}

If both shells are made of dust, one obtains $m_2=0$ and $\hat\sigma_2=-\frac{1}{4\pi r_2^2}$. That surely violates all energy conditions but is interesting in the sense that the curvature is localized just between the two shells and has no effect on the outside world --- similarly to the case of a sandwich plane gravitational wave. However, the exterior shell can easily be made of dust without admitting a negative mass density if a non vanishing field outside ($Q_2\neq 0$) is allowed.

Next, we discuss some properties of the outer shell without any assumption about the inner shell. If the exterior shell is made of dust, though not necessarily the inner shell, then an equation for $q^2$ is implied. The positivity of $q^2$ requires $m_2\leq1$. The case $m_2=1$ leads to $q=0$ and $\hat\sigma_2=0$, which contradicts our definition of a condenser. However, for $m_2<1$ the mass density $\hat\sigma_2$ is negative and all energy conditions are violated. Thus, \emph{the exterior shell cannot be made of dust if any energy condition is to be satisfied}.

Finally we also allow a non vanishing pressure $\hat p_2$. The mass density $\hat\sigma_2$ is positive if either $m_2>1$ or $m_2\leq 1, q^2\geq2 r_2(1-m_2)$. Even though the mass density is positive the Schwarzschild mass outside can again be smaller than between the shells, if the charge is sufficiently large, cf. \eqref{eq:quasilocalmass}. The Schwarzschild mass outside can, in fact, be negative but then the inner shell must have $\hat\sigma_1<0$. Furthermore, for $m_2>0$ all energy conditions can be satisfied for the outer shell with $q^2<1$ (under-extreme case) even if its radius $r_2<r_-$ (i.e., $r_2$ is below $r_-$ which would be the location of the inner horizon corresponding to the inner shell if the outer shell were not there). However, the inner shell violates then at least some energy conditions as discussed above.  To avoid this we may consider just the under-extreme Reissner-Nordstr{\"o}m field instead of the inner shell. Then the ``outer'' shell can satisfy all energy conditions. Interestingly, in this way we construct a spacetime which is Schwarzschild outside the outer (physical) shell inside of which then is just a naked singularity with $M_1^2>Q^2$. Nevertheless, such a configuration cannot be formed by a dynamical process from Cauchy data without a naked singularity.

In the classical regime $r_1,r_2\gg\max[1,m_2]$ the standard classical results \eqref{eq:classical results} are retrieved.

\begin{acknowledgments}

We thank Tom\'a\v s Ledvinka for discussions and the Albert Einstein Institute for kind hospitality. 
JB acknowledges the partial support from Grant No. GACR 202/09/00772 of the Czech Republic, of Grants No. LC06014
and No. MSM0021620860 of the Ministry of Education. NG was financially supported by the Grants No. GAUK 116-10/258025 and No. GACR 205/09/H033.
\end{acknowledgments}

\end{document}